\documentclass[conference]{IEEEtran}
\IEEEoverridecommandlockouts
\usepackage{
amsmath,cite,bm,amsfonts,amssymb,
graphicx,color,mathtools,setspace,
comment,multirow,xfrac,epstopdf,
footnote,multirow,lscape,float,
caption,subcaption,amsthm,alltt,placeins}

\newcommand{\myincludegraphics}
{\includegraphics[trim=0.4cm 0.1cm 1cm 0.8cm, clip=true, width=1\columnwidth]}
\newcommand{\raisecapt}{\vspace{-0.1cm}}
\begin{document}

\bstctlcite{IEEEexample:BSTcontrol} 

\title{3D Mobility Models and Analysis for UAVs}

\author{\IEEEauthorblockN{%
Peter J. Smith\IEEEauthorrefmark{1}, %
Pawel A. Dmochowski\IEEEauthorrefmark{2}, %
Ikram Singh\IEEEauthorrefmark{2}, %
Richard Green\IEEEauthorrefmark{3}, %
Carl P. Dettmann\IEEEauthorrefmark{4}, %
Justin P. Coon\IEEEauthorrefmark{5}}
\IEEEauthorblockA{\IEEEauthorrefmark{1}%
School of Mathematics and Statistics, Victoria University of Wellington, Wellington, New Zealand}
\IEEEauthorblockA{\IEEEauthorrefmark{2}%
School of Engineering and Computer Science, Victoria University of Wellington, Wellington, New Zealand}
\IEEEauthorblockA{\IEEEauthorrefmark{3}%
Department of Computer Science and Software Engineering, University of Canterbury, Christchurch, NZ}
\IEEEauthorblockA{\IEEEauthorrefmark{4}%
School of Mathematics, University of Bristol, Bristol, UK}
\IEEEauthorblockA{\IEEEauthorrefmark{5}%
Department of Enginering Science, University of Oxford, Oxford, UK}
\IEEEauthorblockA{email:%
~\{peter.smith,pawel.dmochowski,ikram.singh\}@ecs.vuw.ac.nz,\\%
~richard.green@canterbury.ac.nz,~carl.dettmann@bris.ac.uk,~justin.coon@eng.ox.ac.uk
}%
}
\maketitle
\begin{abstract}
We present a flexible family of 3D mobility models suitable for unmanned aerial vehicles (UAV). Based on stochastic differential equations, the models offer a unique property of explicitly incorporating the mobility control mechanism and environmental perturbation, while enabling tractable steady state solutions for properties such as position and connectivity. Specifically, motivated by UAV flight data, for a symmetric mobility model with an arbitrary control mechanism, we derive the steady state distribution of the distance from the target position. We provide closed form expressions for the special cases of the Ornstein-Uhlenbeck (OU) process and on-off control (OC). We extend the model to incorporate imperfect positioning and asymmetric control. For a practically relevant scenario of partial symmetry (such as in the x-y plane), we present steady state position results for the OU control. Building on these results, we derive UAV connectivity probability results based on a SNR criterion in a Rayleigh fading environment.  
\end{abstract}
%
%
%
\IEEEpeerreviewmaketitle
%
%
\section{Introduction}
In the modelling of three-dimensional (3D) mobility for mobile devices it is difficult to construct models which are both tractable and general. In broad terms, there are two scenarios of interest. The first, \textit{Scenario 1}, concerns high precision applications where accurate modelling of mobility is required in small volumes. The second, \textit{Scenario 2}, concerns models to provide mobility over wide areas. Obvious examples include the use of unmanned aerial vehicles (UAVs) in high precision agriculture (eg. tree pruning applications \cite{lee2018}) and swarms of UAVs operating over a wide geographical area \cite{DroneCellsHalim}. In the first scenario, it is useful for the model to allow different control mechanisms for the mobile device and to include the effects of imperfect navigation (eg. GPS error). In the second scenario, tractable steady state distributions for position and distance are important as they lead to results on connectivity and signal-to-noise-ratio (SNR) for communication links.

To date a variety of models have been used in the literature for the mobility in ad hoc networks \cite{camp2002survey} and UAV networks \cite{sharmaTWC2019}. Examples include Brownian motion, random direction \cite{nain2005properties,camp2002survey}, random waypoint \cite{bettstetter2003node,hyytia2006spatial} and Gauss-Markov models \cite{Liang2003,camp2002survey}. Recently, to allow for effects of altitude control in addition to spatial excursions, a mixed mobility model has been proposed \cite{sharmaTWC2019} which allows for different behaviour in the horizontal and vertical directions. Each of these synthetic models, while allowing analysis, possesses undesirable features, such as piecewise motion for waypoint models and unbounded wandering in Brownian motion. Critically, they lack the ability to \textit{explicitly} model the control mechanism which attempts to return the node to the desired location\cite{smithcoon}. Hence, in this paper, we are motivated to create a family of 3D mobility models based on stochastic differential equations (SDEs) which explicitly allow the use of different control mechanisms and lead to tractable steady state solutions. The models presented are trivial for system simulations and allow analysis of certain system features such as connectivity and link SNR. We also include the effects of imperfect navigation into the models. In particular the contributions are as follows.
\begin{itemize}
\item For arbitrary, symmetric control, we derive steady state distance distributions, and closed form expressions for the special cases of the OU process and on-off control (OC).
\item To account for effects such as GPS errors and asymmetries in 3D mobility, we extend the model to include imperfect positioning and asymmetric control and perturbation. We derive analytical results for the steady state distance distribution for OU control as well as simple closed form solutions for an important case of partial symmetry (e.g. in the $(x-y)$ plane).
\item Building on these results, we present analytical expressions for the connectivity probability of UAVs based on an SNR criterion in a Rayleigh fading environment.  
\end{itemize}
%

%
\section{Symmetric 3D Mobility Model}\label{Symm}
\begin{figure}[ht]
	\centering
	\includegraphics[trim=0.42cm 0.5cm 0.58cm 0.9cm, clip=true, width=0.60\columnwidth]{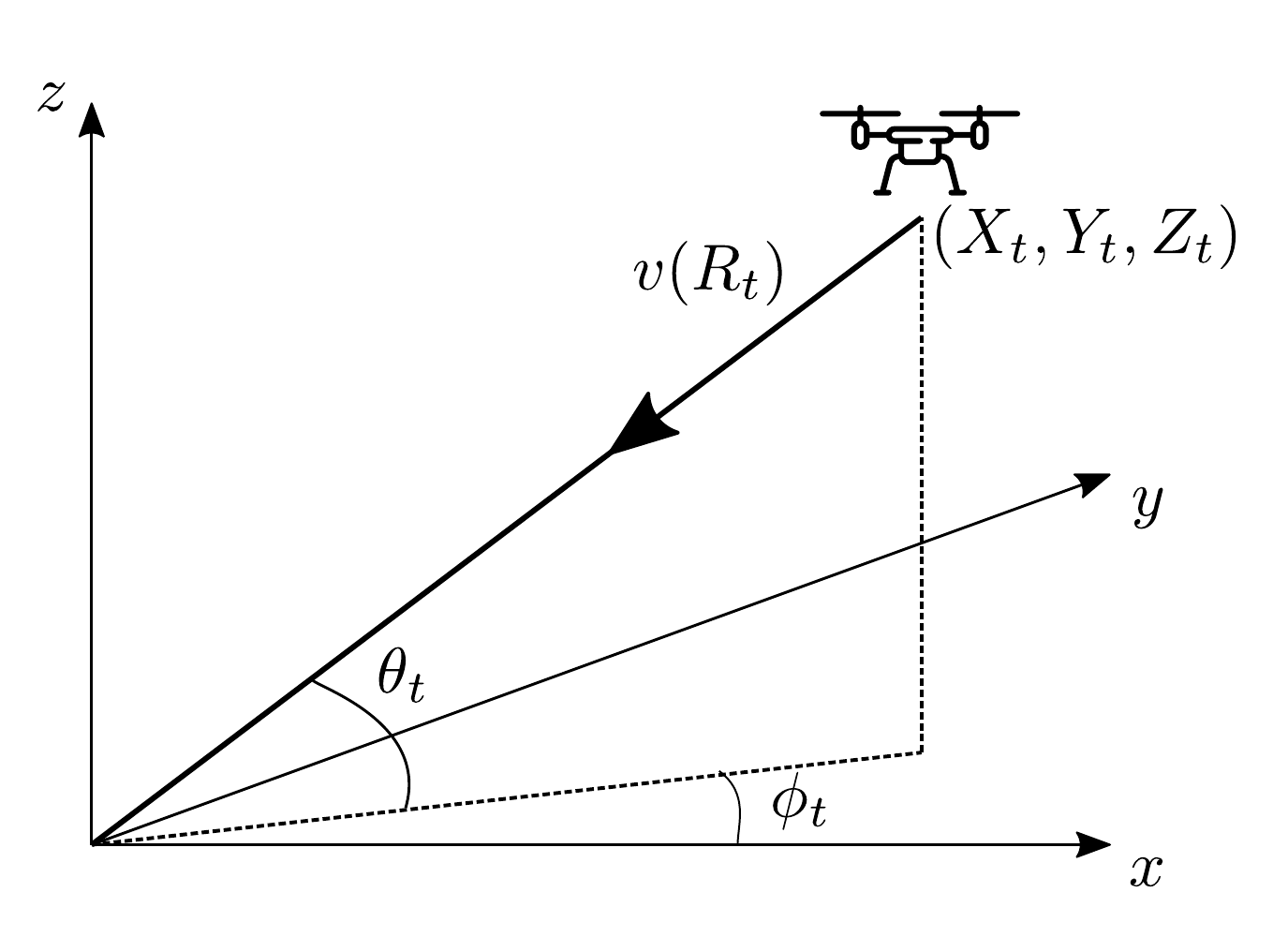}
	\raisecapt\caption{3D model.}
	\label{fig_3Dmodeldiag}
\end{figure}

Consider a device in 3D space located at time $t$ at $(X_t,Y_t,Z_t)$ with radial distance from the origin $R_t=\sqrt{{X_t}^2+{Y_t}^2+{Z_t}^2}$. We assume the on-board control mechanism attempts to maintain position at $(0,0,0)$ by moving the device towards the origin with a velocity $v(R_t)$, which is solely a function of distance. Hence, as shown in Fig.~\ref{fig_3Dmodeldiag}, the control is symmetric in all dimensions - a typical condition for \textit{Scenario 1}.
The device undergoes Brownian perturbations so that the resulting SDEs for position are given by
\begin{align}\label{eq2}
&dX_t=\frac{-v(R_t)}{R_t}X_t{dt}+\sigma{dW}_{1t},\notag\\
&dY_t=\frac{-v(R_t)}{R_t}Y_t{dt}+\sigma{dW}_{2t},\\
&dZ_t=\dfrac{-v(R_t)}{R_t}Z_t{dt}+{\sigma}{dW}_{3t},\notag
\end{align}
where $W_{1t}$, $W_{2t}$ and $W_{3t}$ are three independent standard Brownian motion processes, and $\sigma$ is the perturbation parameter. The Cartesian coordinates are given by $X_t=R_t\cos(\theta_t)\cos(\phi_t)$, $Y_t=R_t\cos(\theta_t)\sin(\phi_t)$, $Z_t=R_t\sin(\theta_t)$ where $\theta_t$ is the angle of elevation $(\theta_t=\sin^{-1}({Z_t}/{R_t}))$ and $\phi_t$ is the azimuth angle $(\phi_t=\tan^{-1}({Y_t}/{X_t}))$. 
%
%
Now, \eqref{eq2} is a standard example of a system of SDEs governed by a multivariate Fokker-Planck (FP) equation \cite[ Eqs. 4.3.41-4.3.42]{gardiner}. Using the Fokker-Planck formulation, in Appendix~\ref{appenA} we show that the steady state PDF of $(X_t,Y_t,Z_t)$ is given by:
\begin{align}
f(x,y,z)=K_0 \exp\left(-\frac{2V(r)}{\sigma^2}  \right),\notag
\end{align}
for some constant, $K_0$, where $r=(x^2+y^2+z^2)^{1/2}$ and $V(r)=\int_{0}^{r} v(\tau) \ d\tau$.

Employing a Cartesian to polar transformation, established methods lead to the steady state PDF of $R_t$,
\begin{align}\label{PDF}
f_R(r)=Kr^2\exp\left({\frac{-2V(r)}{\sigma^2}}\right)\qquad r\ge 0,
\end{align}
where $K$ is a constant ensuring
\begin{align}
\int_0^{\infty}{Kr^2\exp\left({\frac{-2V(r)}{\sigma^2}}\right)dr}=1.
\end{align}
%
%
\subsection{Special Cases}\label{secPerfNavSpecial}
Here, we consider two useful special cases: the OU process where $v({R_t}) \propto {R_t}$ and on-off control (OC), where $v(\cdot)$ is either ON (constant velocity) or OFF (no control).\\
{\bf{OU}}: Here, $v({R_t}) =\alpha{R_t}$, so that $V(\tau)=\alpha \tau^2/2$ and (\ref{PDF}) becomes
\begin{align}\label{PDF_OU}
f_R(r)=\frac{4\alpha^{3/2}}{\sqrt{\pi}\sigma^3}r^2\exp\left({\frac{-\alpha r^2}{\sigma^2}}\right)\qquad r\ge 0.
\end{align}
This leads by simple integration to the CDF:
\begin{align}\label{CDF_OU}
F_R(r)=   {\textrm{erf}}  \left(\sqrt{\frac{\alpha r^2}{\sigma^2}}\right)-\sqrt{\frac{4\alpha r^2}{\pi \sigma^2}}\exp(-\alpha r^2/\sigma^2)          \quad r\ge 0,
\end{align}
where ${\textrm{erf}}(\cdot) $ is the error function.

{\bf{OC}}: Here, $v({R_t}) =c$ if $R_t>m$ and $v({R_t}) =0$ otherwise. Hence, the control is ON at constant velocity $c$ when the displacement exceeds the threshold $m$ and is OFF (zero) otherwise. Hence, $V(\tau)=c(\tau-m)$ for $\tau>m$ and is zero otherwise. Substituting into (\ref{PDF}) and integrating gives the CDF
%
\begin{equation}\label{CDF_OC}
F_R(r) =	Kr^3/3,  \qquad  r \le m 
\end{equation}
and 
\begin{align}\label{CDF_OC2}
F&_R(r) =	\frac{Km^3}{3}+
\frac{K \sigma^4}{4 c^3}
\bigg\{ 	\sigma^4+2mc\sigma^2+2m^2c^2   
\bigg.  \\ 
&\bigg. -\left( \sigma^4+2rc\sigma^2+2r^2c^2 \right)
       \exp\left(-\frac{2c(r-m)}{\sigma^2}\right) 
\bigg\}, \ r \ge m,\notag
\end{align}
where $K=12c^3[3\sigma^2(\sigma^4+2mc\sigma^2+2m^2c^2)+4m^3c^3]^{-1}$.
Note that (\ref{PDF}) is a completely general solution for the steady state distance of a device from the origin, with an arbitrary, radially dependent (symmetric) control mechanism. In Sec.~\ref{NumRes} we show that symmetric control can be a reasonable model for UAVs in high precision applications. Note that (\ref{PDF}) is given in closed form, except for the constant $K$ and the function $V(r)$ which are defined as integrals. For simple control mechanisms, as for OU and OC, $f_R(r)$ and $F_R(r)$ are available in closed form. Similarly any piecewise linear control function can be solved.
%
%
%
\section{Mobility models with imperfect positioning}\label{asymm}
Consider the case where only imperfect position information is available at the device - a practical consideration for both \textit{Scenario 1} and \textit{Scenario 2}. We also extend the mobility model to include asymmetries in both control and perturbation. This is motivated by the observation that although symmetry in the $x-y$ plane is a reasonable model, it is certainly possible for motion in the $z$ direction to behave differently, in particular for the large scale deployment of \textit{Scenario 2}. At time $t$, the true position is $(X_t,Y_t,Z_t)$ but only the errored coordinates, $(\hat{X}_t,\hat{Y}_t,\hat{Z}_t)$ are available where  $\hat{X}_t=X_t+\epsilon_{1t}$, $\hat{Y}_t=Y_t+\epsilon_{2t}$, $\hat{Z}_t=Z_t+\epsilon_{3t}$ and $\hat{R}_t=\sqrt{{\hat{X}_t}^2+{\hat{Y}_t}^2+{\hat{Z}_t}^2}$. The errors are modelled so that they are Gaussian and smoothly varying with no drift. The OU model is a convenient approach to  providing these properties. Hence, the error terms are defined by \cite{gardiner}
\begin{align}\label{pert_process}
d\epsilon_{it}=-\beta_i\epsilon_{it}dt+s_idB_{it} \quad \textrm{for } i=1,2,3,
\end{align}
where the initial error values are zero ($\epsilon_{it}=0$ for $t=0$) and $B_{it} $ are iid standard Brownian motion processes. The positive parameters, $\beta_i$ and $s_i$, control the variability of the errors (possibly different in each dimension) and the steady state distribution of the errors is $\epsilon_{it}\sim\mathcal{N}(0,{s_i^2}/{2\beta_i})$.
With these positioning errors, the control mechanism in (\ref{eq2}) is based on the estimated positions so that the SDEs extended to the non-symmetric case become:
\begin{align}\label{imperfectSDE}
&dX_t=\frac{-{v_1}(\hat{R}_t)}{\hat{R}_t}\hat{X}_t{dt}+\sigma_1{dW}_{1t},\notag\\
&dY_t=\frac{-{v_2}(\hat{R}_t)}{\hat{R}_t}\hat{Y}_t{dt}+\sigma_2{dW}_{2t},\\
&dZ_t=\frac{-{v_3}(\hat{R}_t)}{\hat{R}_t}\hat{Z}_t{dt}+{\sigma_3}{dW}_{3t},\notag
\end{align}
with the error process given by (\ref{pert_process}). To the best of our knowledge there is no tractable solution for the system of SDEs given by (\ref{pert_process}) and (\ref{imperfectSDE}) even for the symmetric case. Hence, in order to make analytical progress, we consider the classical  OU process in 3D where $v_i({\textrm{displacement}}) \propto {\textrm{displacement}}$. This is described by
\begin{align}\label{eq3DOU}
&dX_t=-\alpha_1\hat{X}_t{dt}+\sigma_1{dW}_{1t},\notag\\
&dY_t=-\alpha_2\hat{Y}_t{dt}+\sigma_2{dW}_{2t},\\
&dZ_t=-\alpha_3\hat{Z}_t{dt}+{\sigma_3}{dW}_{3t},\notag
\end{align}
with the error process given by (\ref{pert_process}). This 6-dimensional process is separable into three two-dimensional processes, $(X_t,\epsilon_{1t})$,  $(Y_t,\epsilon_{2t})$ and  $(Z_t,\epsilon_{3t})$. Each 2D process is a multivariate OU process and it is shown in Appendix~\ref{appenB} that the resulting steady state distributions are $X_t\sim\mathcal{N}(0,\lambda_1)$,  $Y_t\sim\mathcal{N}(0,\lambda_2)$ and  $Z_t\sim\mathcal{N}(0,\lambda_3)$ where 
\begin{align}\label{lambdai}
\lambda_i=\frac{\sigma_i^2}{2  \alpha_i}+\frac{\alpha_i }{(\alpha_i+\beta_i)}\frac{s_i^2}{2  \beta_i}.
\end{align}
This solution clearly demonstrates the effect of position error and as an example can be rewritten for $X_t$ as
\begin{align}
{\textrm{Var}}(X_t)={\textrm{Var}}(X_t^{\textrm{error free}})+\frac{1 }{1+\beta_1/\alpha_1}{\textrm{Var}}(\epsilon_{1t}).
\end{align}
Hence, ${\textrm{Var}}(X_t)$ ranges form its ideal value, $ {\textrm{Var}}(X_t^{\textrm{error free}})$ to its upper limit $ {\textrm{Var}}(X_t^{\textrm{error free}})+{\textrm{Var}}(\epsilon_{1t})$ as the ratio of the control parameters, $\beta_1/\alpha_1$, changes. 

Note that this solution allows different control parameters, $\alpha_i$, in each dimension, as well as different levels of perturbation, different $\sigma_i$. Similarly the error processes are different in each dimension. However, the price to be paid for this generality is that the control mechanism is the simple one where $v_i({\textrm{displacement}}) \propto {\textrm{displacement}}$ in each dimension. 

{\bf{Unequal ${\mathbf{\lambda}}_i$} values}: In the general case when the three values of $\lambda_i$ are all different then $R_t^2$ is a quadratic form  in Gaussian random variables with the formulation  $R_t^2=\lambda_1 W_1^2+\lambda_2 W_2^2+\lambda_3 W_3^2$, where $W_1$, $W_2$ and $W_3$ are iid $\mathcal{N}(0,1)$ variables. Hence,  $R_t^2$ has a known CDF \cite[pp. 156]{JandK}. This immediately gives the steady state CDF of $R_t$ as:
\begin{align}\label{asymmetric}
F_R(r)=P(R_t \le r)=\sum_{j=0}^{\infty}e_jP(\chi^2_{3+2j} \le r^2/\eta),
\end{align}
where $\chi^2_{k}$ is a Chi-squared random variable with $k$ degrees of freedom, $\eta$ is an arbitrary constant and 
$e_s$ is given by
\begin{align}
\label{asymmetric_e}
e_s=\frac{1}{2s}\sum_{j=0}^{s-1}H_{s-j}e_j, \quad e_0=\sqrt{\frac{\eta^3}{\lambda_1 \lambda_2 \lambda_3}},
\end{align}
where $H_s=\sum_{j=1}^3(1-\eta/\lambda_j)^s$. We adopt the typical choice of $\eta$ as $\eta=3[1/\lambda_1+1/\lambda_2+1/\lambda_3]^{-1}$ \cite{JandK} and note that the Chi-squared CDFs required in (\ref{asymmetric}) are known functions which can be expressed as finite sums or incomplete gamma functions.

{\bf{Some equal ${\mathbf{\lambda}}_i$ values}}: In the symmetric case where all ${\mathbf{\lambda}}_i$ values are equal, we default back to Sec.~\ref{Symm}. Here, $R_t^2=\lambda_1 \chi^2_3$ and all results are known based on known properties of the Chi-squared  variable. The more interesting case is $\lambda_1=\lambda_2 \ne \lambda_3$ where the control and errors are symmetric in the $x-y$ plane but different in the $z$ direction. Letting $U=(W_1^2+ W_2^2)/2$ and noting that $U$ is exponential with unit mean, we have
\begin{align}\label{equal2}
F_R(r)&=P(2\lambda_1 U +\lambda_3 W_3^2 \le r^2), \\
&=\mathbb{E}\left[1_{W_3<r/\sqrt{\lambda_3}}\left(1-\exp\left(\frac{\lambda_3W_3^2-r^2}{2\lambda_1}\right)\right)\right],\notag					
\end{align}
where $1_{x \in A}$ is the indicator function that equals 1 if ${x \in A}$ and is zero otherwise. Next, using the fact that $V=W_3^2 \sim \chi^2_1$, we have the PDF of $V$ which allows (\ref{equal2}) to be written as
\begin{align}\label{equal2b}
F_R(r)&=\int_0^{r^2/\lambda_3}\left[1-\exp\left(\frac{\lambda_3v-r^2}{2\lambda_1}\right)\right]\frac{\exp(-v/2)}{2\sqrt{2\pi v}}dv.\notag \\					
\end{align}
Using \cite[Eq.3.361.1 and 8.252.1]{gradshteyn}, the CDF can be solved for the two cases, $\lambda_1>\lambda_3$ and  $\lambda_1<\lambda_3$ as:
\begin{align}\label{equal2c}
F_R(r)= {\textrm{erf}}\left(  \sqrt{\tfrac{r^2}{2 \lambda_3}}\right)-{\frac{\sqrt{\lambda_1}e^{-r^2/(2\lambda_1)}}{\sqrt{\lambda_1-\lambda_3}}}{\textrm{erf}}\left( \sqrt{\tfrac{r^2(\lambda_1-\lambda_3)}{2\lambda_1\lambda_3}}\right)					
\end{align}
and 
\begin{align}\label{equal2d}
F_R(r)&= {\textrm{erf}}\left(  \sqrt{\frac{r^2}{2 \lambda_3}}\right)-{\frac{\sqrt{\lambda_1}e^{-r^2/(2\lambda_1)}}{\sqrt{\lambda_3-\lambda_1}}}{\textrm{erfi}}\left( \sqrt{\tfrac{r^2(\lambda_3-\lambda_1)}{2\lambda_1\lambda_3}}\right)				
\end{align}
respectively.
%
%
\section{Connectivity}\label{ImpGen}
We now examine the probability of connectivity for mobile devices. Consider a link of distance, $R_t$, at time $t$ in a Rayleigh fading environment with path loss exponent, $\gamma$. In the absence of shadowing, the SNR of a single input single output (SISO) link is ${\textrm{SNR}}_t=AR_t^{-\gamma}|h_t|^2$, where $|h_t|^2 \sim {\textrm{Exp}}(1)$ and $A$ is  a constant accounting for transmit power, receiver noise, etc. If connectivity relies on the SNR exceeding a threshold, ${\textrm{SNR}}_0$, then the probability of connectivity, $P_{\textrm{conn}}$, is given by
\begin{align}\label{pconn}
P_{\textrm{conn}}&=	P\left(AR_t^{-\gamma}|h_t|^2>{\textrm{SNR}_0}\right)	\notag\\
&=E\left[P\left(R_t<(A|h_t|^2/{\textrm{SNR}_0})^{1/\gamma}\,\big|\, |h_t|^2\right)	\right]		\notag\\
&=\int_0^{\infty} F_R(Bx^{1/\gamma})e^{-x}dx,
\end{align}	
where $B=(A/{\textrm{SNR}_0})^{1/\gamma}$. Note that (\ref{pconn}) assumes that steady state has been reached and $F_R(r)$ is the steady state CDF of $R_t$. Computationally, for any value of $\gamma$, (\ref{pconn}) can be evaluated via simple numerical integration as the integrand is smooth and exponentially decaying in the upper tail. However, for the {\textit{edge cases}} of $\gamma=2$ and $\gamma=4$, so-called because $2 \le \gamma \le 4$ is the usual range of values for the path loss exponent, analytical progress using (\ref{pconn}) is possible. Two examples are given below.
%
%
%
\subsection{Symmetric Mobility Models}
The work in Sec.~\ref{Symm} mainly focused on high precision applications where connectivity is not normally a problem. However, if the model in (\ref{eq2}) is applied to environments where outage is a factor, then a general solution is given by substituting (\ref{PDF}) into (\ref{pconn}) which gives
\begin{align}\label{pconn_sec2}
P_{\textrm{conn}}=\int_0^{\infty}e^{-x}\int_0^{u(x)}Kr^2\exp\left(\frac{-2V(r)}{2}\right)drdx,
\end{align}
where $u(x)=Bx^{1/\gamma}$ and $B=({\textrm{SNR}_0}/A)^{1/\gamma}$. More directly, if the CDF, $F_R(r)$, is known then (\ref{pconn}) can be used directly. 

As an example, consider the non-linear control mechanism, OC, and $\gamma=2$. Here, substituting (\ref{CDF_OC}) and (\ref{CDF_OC2}) into (\ref{pconn}) gives
\begin{align}\label{pconn_OC}
&P_{\textrm{conn}}=\frac{1}{2}\int_0^{(m/B)^2}{KB^2x}e^{-x}dx+\int_{(m/B)^2}^{\infty}e^{-x} \times\notag\\
& \left(1-\frac{K\sigma^4}{4c^2}\left(1+\frac{2cB\sqrt{x}}{\sigma^2}\right)\exp\left(-2c(B\sqrt{x}-m)/\sigma^2\right)\right)dx.
\end{align}
In (\ref{pconn_OC}), the integrals $\int x\exp(-x)dx$ and $\int\exp(-x)dx$ are trivial. The remaining integrals $\int\exp(-p\sqrt{x}-x)dx$ and $\int\ \sqrt{x}exp(-p\sqrt{x}-x)dx$ can be solved using the substitution $v=\sqrt{x}$, followed by integration by parts and the use of \cite[Eq. 3.322.1]{gradshteyn}. This gives:
\begin{align}\label{pconn_OC_final}
P_{\textrm{conn}} & =\frac{K}{2}  
           \left[ 
					     B^2-(m^2+B^2) e^{-\left(\frac{m}{B}\right)^2 }   
					 \right] 
					   +e^{ -\left(\frac{m}{B}\right)^2 }
																 \\
									& -\frac{K\sigma^4}{4c^2}e^{\frac{2cm}{\sigma^2}}
									\left\{
									     \left(  
											    1+\frac{2cB}{\sigma^2}    
									        \left(  \frac{m}{B}-\frac{cB}{\sigma^2}  \right)   
									     \right)  
											 e^{ -\frac{2cm}{\sigma^2}-\left(\frac{m}{B} \right)^2  }
									\right. 
									\notag\\									
							    & 
									\left.
								\left[ 
												1-\textrm{erf} \left( 
												       \frac{cB}{\sigma^2} +\frac{m}{B}  
												      \right) 
								\right]	
								\frac{\sqrt{\pi}}{2}	\frac{4c^3B^3}{\sigma^6}
								 e^{ \frac{c^2 B^2}{\sigma^4}  }																
									\right\} .
									\notag
\end{align}
%
%
%
\subsection{Non-symmetric Mobility Models}
More importantly, we consider the connectivity probability for the models in Sec.~\ref{asymm} which are designed to apply in large scale applications. In the most general case where there is asymmetry in all three dimensions, (\ref{asymmetric}) applies and
\begin{align}\label{pconn_asymm3}
&P_{\textrm{conn}}=\sum_{j=0}^{\infty}e_j\int_0^{\infty}P\left(\chi^2_{3+2j} \le \frac{B^2x^{2/\gamma}}{\eta}\right)e^{-x}dx.\notag\\
\end{align}
For a Chi-squared random variable, the CDF is an incomplete gamma function and (\ref{pconn_asymm3}) becomes
\begin{align}\label{pconn_asymm4}
&P_{\textrm{conn}}=\sum_{j=0}^{\infty}\frac{e_j}{\Gamma(j+3/2)}\int_0^{\infty}\gamma\left(j+3/2,\frac{B^2x^{2/\gamma}}{2\eta}\right)e^{-x}dx.\notag\\
\end{align}
For $\gamma=2$, \cite[Eq.6.451.1]{gradshteyn} gives
\begin{align}\label{pconn_asymm5}
P_{\textrm{conn}}=\sum_{j=0}^{\infty}{e_j}\left(1+2\eta/B^2\right)^{-(j+3/2)}.
\end{align}
For $\gamma=4$, we use \cite[Eq. 6.454]{gradshteyn} to give
\begin{align}\label{pconn_asymm6}
P_{\textrm{conn}}{=}\sum_{j=0}^{\infty}{e_j}\left(\frac{B^2}{2 \eta \sqrt{2}}\right)^{j{+}\frac{3}{2}}\exp\left(\frac{B^4}{32\eta^2}\right)D_{{-}j{-}\frac{3}{2}}\left(\frac{B^2}{2 \eta \sqrt{2}}\right),
\end{align}
where $D_{\nu}(\cdot)$ is the parabolic cylinder function \cite[Sec.9.24]{gradshteyn}.

In the useful scenario, where there is symmetry in the $x,y$ plane and the movement in the $z$ direction is more limited, then (\ref{equal2c}) holds and 

\begin{align}\label{pconnasymm}
&P_{\textrm{conn}}= \int_0^{\infty}{\textrm{erf}}\left(  \sqrt{\tfrac{B^2x^{2/\gamma}}{2 \lambda_3}}\right)e^{-x}dx\\
&-\int_0^{\infty}{\frac{\sqrt{\lambda_1}e^{-B^2x^{2/\gamma}/(2\lambda_1)}}{\sqrt{\lambda_1-\lambda_3}}}{\textrm{erf}}\left( \sqrt{\tfrac{B^2x^{2/\gamma}(\lambda_1-\lambda_3)}{2\lambda_1\lambda_3}}\right)	e^{-x}dx.				\notag
\end{align}

For $\gamma=2$, using \cite[Eq. 6.283.2]{gradshteyn} we have 
\begin{align}\label{pconnasymm2}
&P_{\textrm{conn}}= \sqrt{\frac{B^2}{B^2+2\lambda_3}}
\left[  
  1-\frac{2\lambda_1}{B^2+2\lambda_1}\sqrt{\frac{\lambda_1-\lambda_3}{\lambda_1}}
\right].				
\end{align}
For $\gamma=4$, the second integral in \eqref{pconnasymm} is unavailable in closed form and a series expansion is required. Details are not provided for reasons of space.

%
%
\section{Numerical results}\label{NumRes}
\subsection{UAV flight data set}
\begin{figure}[ht]
	\centering
	\myincludegraphics{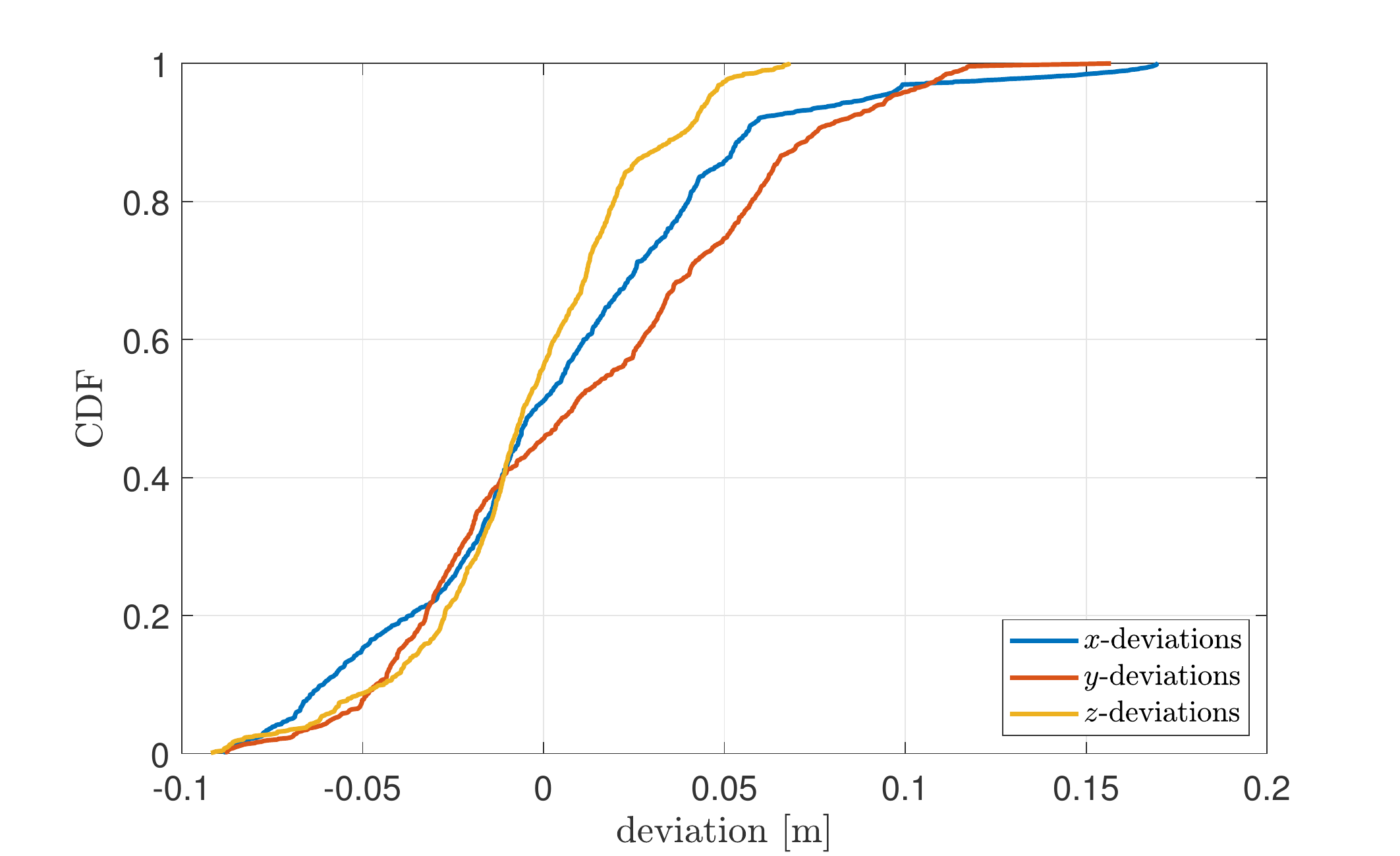}
	\raisecapt\caption{Empirical CDFs of $x, y, z$ values relative to UAV target position.}
	\label{fig_UAVxyzCDF}
\end{figure}

We begin by looking at an example dataset collected from an indoor UAV flight in the University of Canterbury Drone Lab. The quad-rotor UAV has a custom-built air-frame with a pruning arm for agricultural applications and a total weight of 5 kg. The main platform is 900 mm long and the diagonal centre-prop to centre-prop distance is 600 mm. The on-board control mechanism is a Pixhawk flight controller running PX4. The position of the UAV is measured using camera vision techniques giving positional information which is accurate to 1 mm for close range operation (within 1 m) \cite{schofield2018}. The 3D positional data during hovering is measured relative to the target position, so the $x,y,z$ coordinates are errors from the desired location. Fig.~\ref{fig_UAVxyzCDF} shows the empirical CDFs of the $x,y,z$ values. Clearly there are differences for this single flight in the 3 dimensions, but these differences are relatively small. This is an example of \textit{Scenario 1}, where the symmetric model in Sec.~\ref{Symm} is applicable. 
While not shown due to space limitations, examining the vertical changes in the $z$ direction against the distance from the target in the $x,y$ plane, one notes a negligible correlation of -0.076. This indicates that the control is essentially separate in each dimension and thus the separable models in Sec.~\ref{ImpGen} could be applied.

%
%
\subsection{Modelling Results}
\begin{figure}[ht]
	\centering
	\myincludegraphics{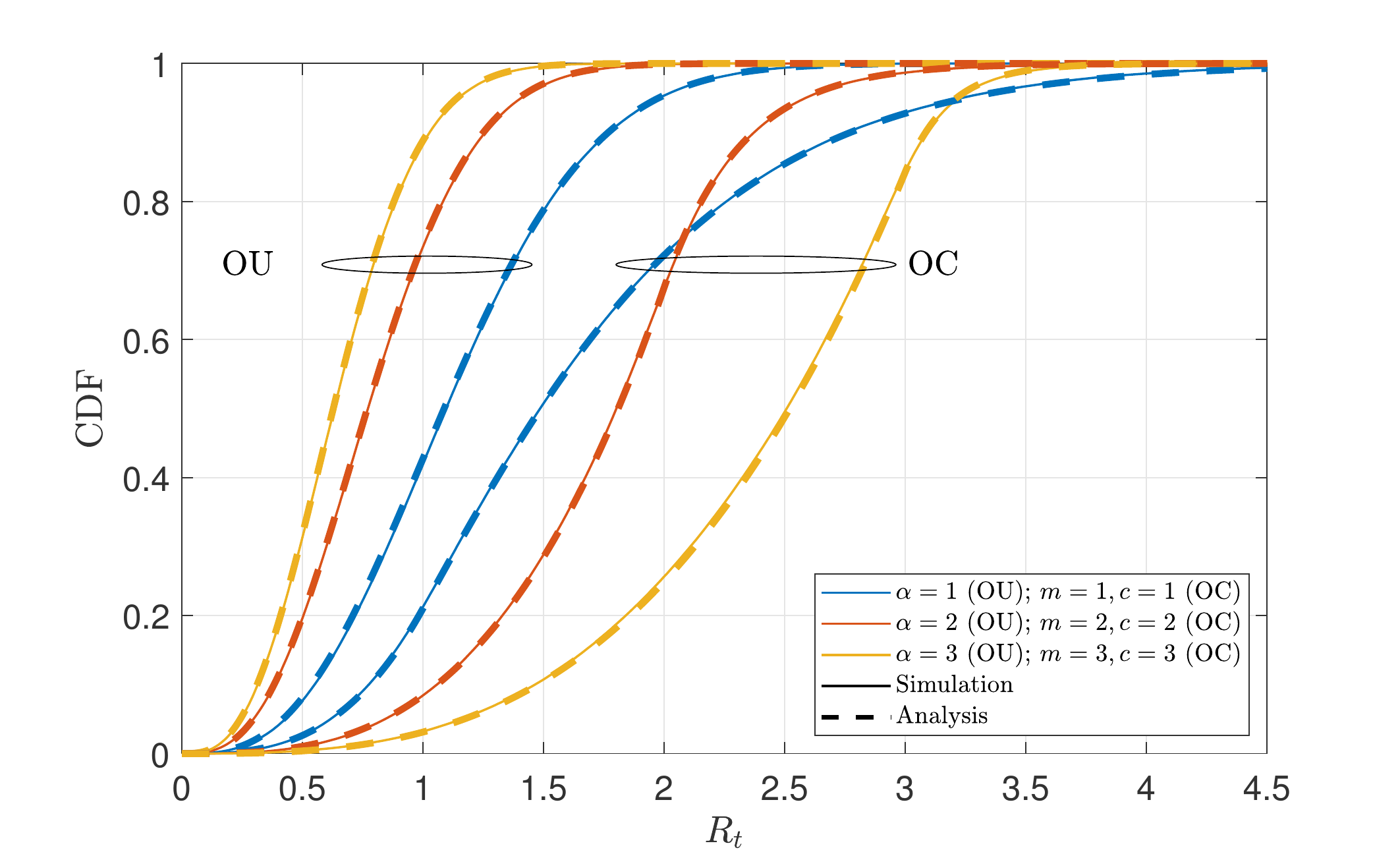}
	\raisecapt\caption{Distance from target CDF for OC ($c=1$, $m=1$) and OU ($\alpha=1$) control; perfect navigation.}
	\label{fig_OCOU_perfNav_CDF}
\end{figure}

We begin with a symmetric case typical of \textit{Scenario 1}. Fig. \ref{fig_OCOU_perfNav_CDF} shows the simulated and analytical CDFs of the distance from the target position for OC and OU control mechanisms for perfect navigation. For OC, the on velocity $c=1$ and the radial distance threshold $m=1$ while for OU the velocity scaling factor $\alpha=1$. The analytical results were computed using \eqref{CDF_OU} for OU and \eqref{CDF_OC}-\eqref{CDF_OC2} for OC. 
As expected, in the case of OU control, increasing the control velocity $\alpha$ yields smaller steady state distance-to-target, as does reducing the distance threshold for OC. For the OC parameters considered, reducing the distance threshold $m$ is coupled with reducing the velocity $c$, and thus the trend reverses at the high CDF tail. 

\begin{figure}[ht]
	\centering
	\myincludegraphics{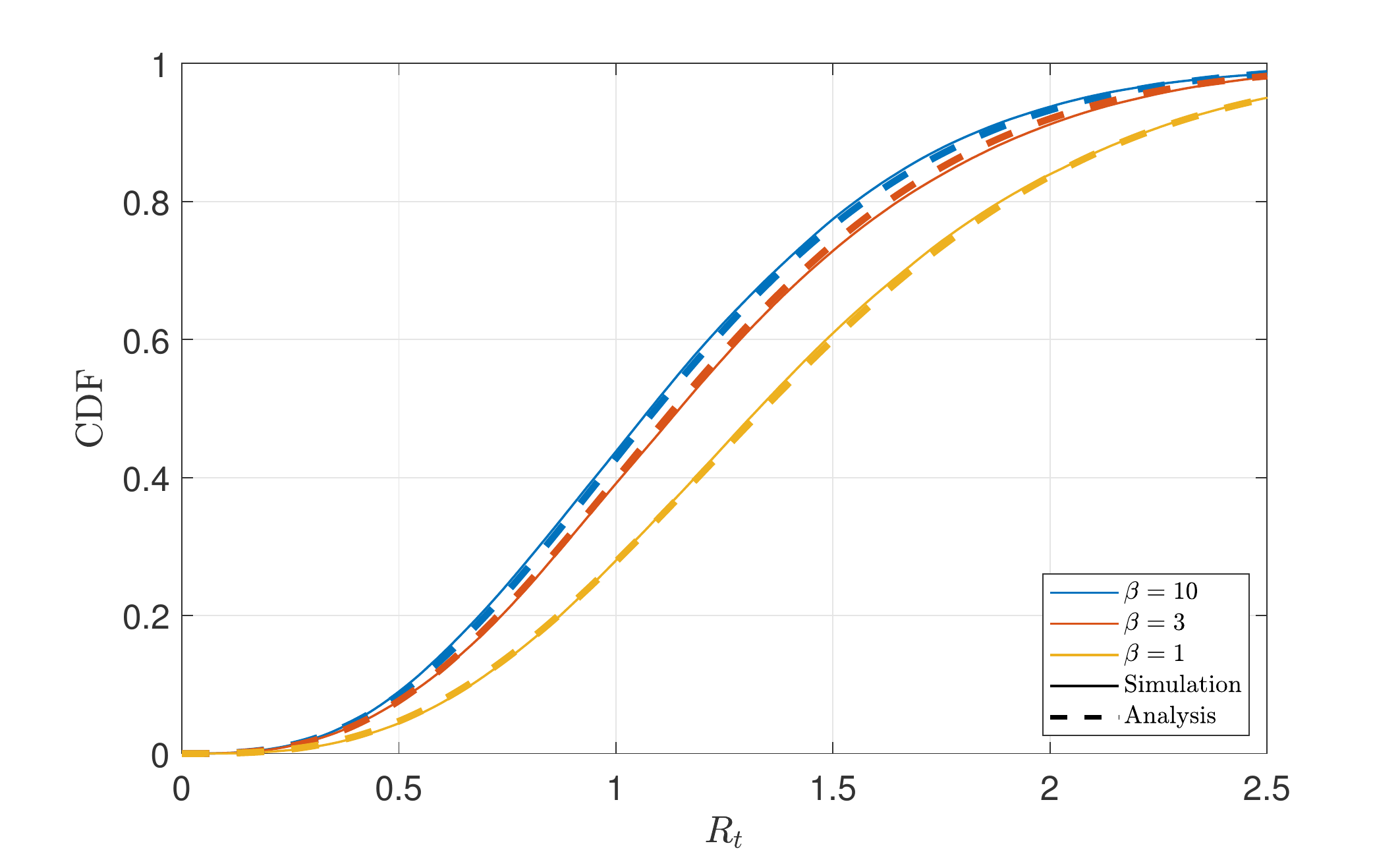}
	\raisecapt\caption{Distance from target CDF for OU control; imperfect navigation ($\alpha_i=1, s_i=1$, $\sigma_1=1.3$, $\sigma_2=1$, $\sigma_3=0.7$).}
	\label{fig_OU_imperfNav_CDF}
\end{figure}

Fig. \ref{fig_OU_imperfNav_CDF} shows the simulated and analytical CDFs of the distance from the target position for the OU control mechanism for imperfect navigation with unequal $\lambda_i$ values, computed using \eqref{lambdai} with $s_i=1$, $\alpha_i=1$, $\beta_i=1, 3, 10$, $\forall i$ and $\sigma_1=1.3$, $\sigma_2=1$ and $\sigma_3=0.7$. The analytical expressions are computed using \eqref{asymmetric} and \eqref{asymmetric_e}.
Predictably, we note that reducing the variability of positioning errors (i.e., increasing $\beta_i$) improves the target position of the UAV, with the performance improvement diminishing for $\beta_i >3$. 

\begin{figure}[ht]
	\centering
	\myincludegraphics{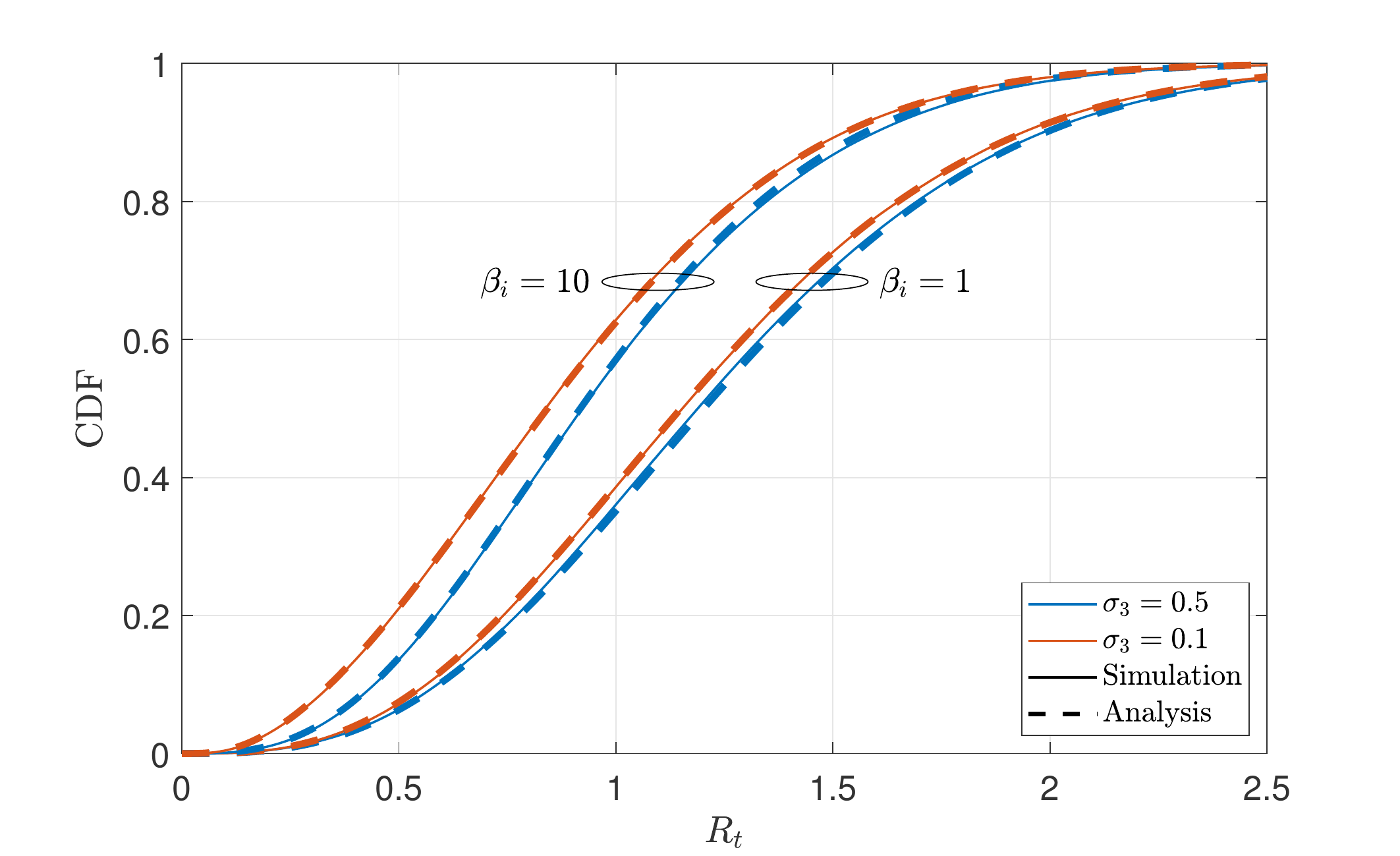}
	\raisecapt\caption{Distance from target CDF for OU control with imperfect navigation ($\alpha_i=1, s_i=1$, $\sigma_1=\sigma_2=1$).}
	\label{fig_OU_imperfNavSomeEqual_CDF}
\end{figure}

Fig. \ref{fig_OU_imperfNavSomeEqual_CDF} shows equivalent results for the case $\lambda_1=\lambda_2\neq\lambda_3$. Here we consider a case where the $z$-direction errors are smaller than those in the $x-y$ plane. Specifically, we let $\sigma_1=\sigma_2=1$ and consider cases for $\sigma_3=0.5, 0.1$ and the navigation errors with $\beta_i=1, 10$ $\forall i$. Here, analytical results are computed using \eqref{equal2c}.  
As expected, we note the improvement in steady state position accuracy with decreasing positioning errors and the vertical perturbation $\sigma_3$. 

\begin{figure}[ht]
	\centering
	\myincludegraphics{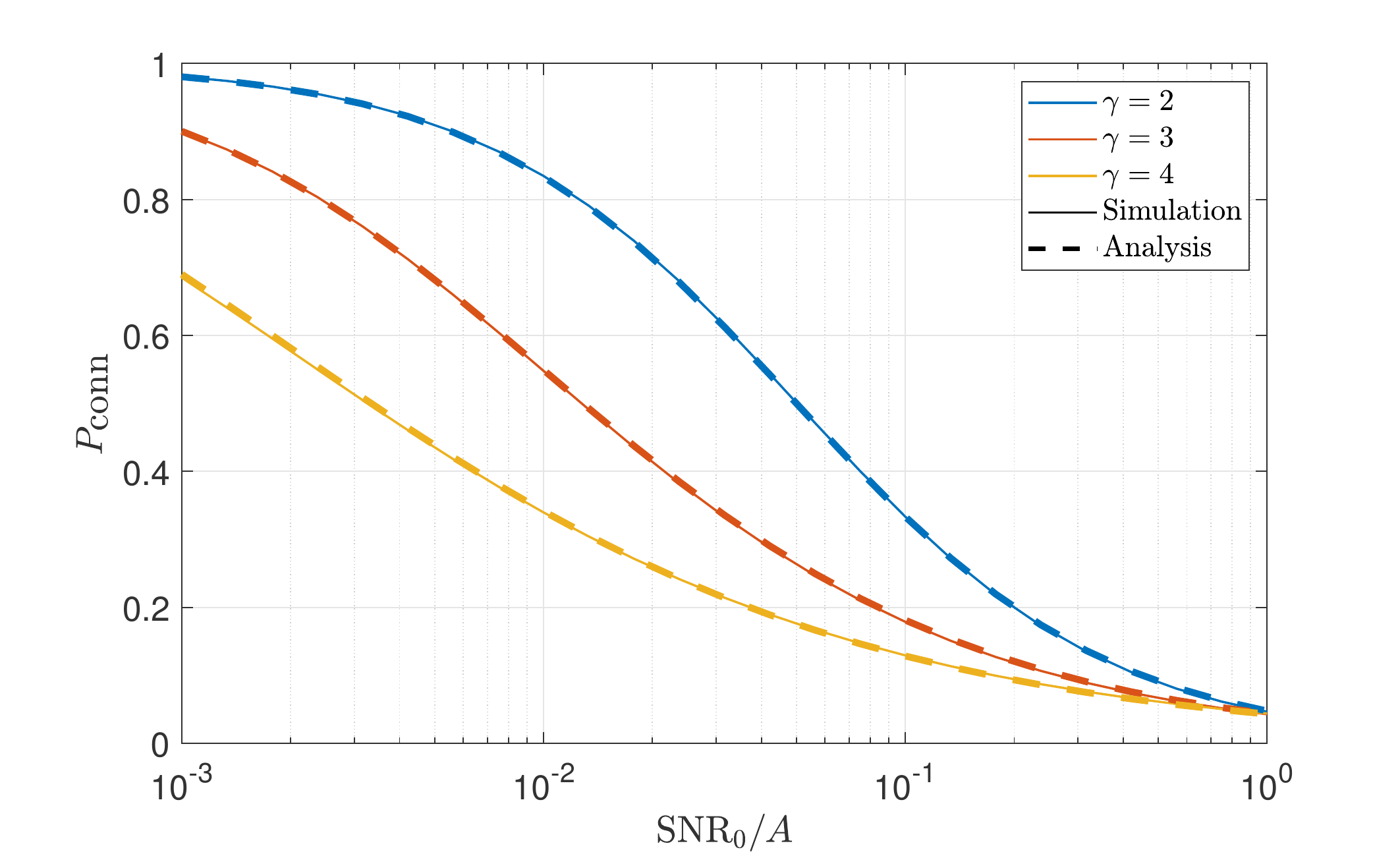}
	\raisecapt\caption{Connectivity probability vs SNR threshold $\textrm{SNR}_0/A$ ($\alpha_i=1, s_i=1$, $\sigma_1=\sigma_2=1$, $\sigma_3=0.01$, $\beta_i=10$).}
	\label{fig_OU_Pconn}
\end{figure}

Finally, we examine the UAV connectivity behaviour in a \textit{Scenario 2} deployment with $\beta_i=10$ and a highly asymmetric values $\sigma_1=\sigma_2=1$, $\sigma_3=0.01$. Fig. \ref{fig_OU_Pconn} shows the simulated and analytical connectivity probability as a function of the threshold $\textrm{SNR}_0/A$. We consider the pathloss exponents of $\gamma=2$ (with analytical results computed using \eqref{pconnasymm2}) and $\gamma=3,4$ (with analytical results computed using \eqref{pconnasymm}).
We note the effects of the pathloss exponent on the connectivity probability, with a dramatic reduction from $\gamma=2$ to $\gamma=4$. Noting the logarithmic x-axis, we see a sharp reduction in connectivity with increasing SNR threshold.
%
%
%
\section{Conclusion}\label{con}
3D mobility models based on stochastic differential equations, incorporating the mobility control mechanism, were presented. Motivated by UAV flight data, for a symmetric mobility model, we derived steady state distance distributions for arbitrary control including closed form expressions for the special cases of OU and OC control. The model was extended to imperfect positioning and asymmetric control. For an important scenario of partial symmetry (in the x-y plane), we presented steady state position results for the OU control and subsequently derived UAV connectivity probability results based on a SNR
criterion in a Rayleigh fading environment.
%
%
\begin{appendices}
\section{Symmetric steady state distribution}\label{appenA}
The drift term in the FP equation is the 3D vector:
\begin{equation}
\mathbf{A}(X_t,Y_t,Z_t)=-\left(\frac{v(R_t)}{R_t}\right)(X_t,Y_t,Z_t)^T
\end{equation}
and the diffusion term is $\mathbf{B}=\sigma^2\mathbf{I}_3$ \cite[pp. 94-95]{gardiner}. From \cite[Eq. 6.2.7-6.2.11, p.141]{gardiner}, we see that the steady state distribution of the process in \eqref{eq2} depends on the functions
\begin{align}\label{eq3}
Z_i(X_t,Y_t,Z_t)&=\sum_{k =1}^3B^{-1}_{ik} \bigg( 2A_k(X_t,Y_t,Z_t) \notag\\
&-\left. \frac{\partial B_{k1}}{\partial{X_t}}-\frac{\partial B_{k2}}{\partial{Y_t}} -\frac{\partial B_{k3}}{\partial{Z_t}}\right),
\end{align}
for $i=1,2,3$ where $\mathbf{B}=(B_{ij})$ and $A_k(X_t,Y_t,Z_t)$ is the $k^{th}$ element of $\mathbf{A}(X_t,Y_t,Z_t)$. Substituting $\mathbf{A}(X_t,Y_t,Z_t)$ and $\mathbf{B}$ into \eqref{eq3} gives
\begin{align}
Z_i(X_t,Y_t,Z_t)&=\frac{2A_i(X_t,Y_t,Z_t)}{\sigma^2}\qquad\textrm{for} \, i=1,2,3.\notag
\end{align}
From \cite[Sec. 6.2.2]{gardiner} the steady state distribution $f(x,y,z)$ for $(X_t,Y_t,Z_t)$ is given by
the solution of 
\begin{equation}\label{partlogf}
\frac{\partial}{\partial{a_i}}\log\left\{f(a_1, a_2, a_3)\right\}
=Z_i(a_1, a_2, a_3)
\end{equation}
%
%
as long as the condition 
\begin{equation}
\frac{\partial{Z_i(x,y,z)}}{\partial{a_j}}=\frac{\partial{Z_j(x,y,z)}}{\partial{a_i}}
\end{equation}
is satisfied where $(a_1,a_2,a_3)=(x,y,z)$. Consider the example $i=1,j=2$. Defining $r=\sqrt{x^2+y^2+z^2}$ we have
\begin{align}
\dfrac{\partial{Z_1(x,y,z)}}{\partial{y}}=\dfrac{\partial (-xv(r)/r)}{\partial{y}}=\dfrac{xy}{r^3}\bigg(v(r)-rv'(r)\bigg),\notag
\end{align}
and
\begin{align}
\dfrac{\partial{Z_2(x,y,z)}}{\partial{x}}=\dfrac{\partial (-yv(r)/r )}{\partial{x}}=\dfrac{xy}{r^3}\bigg(v(r)-rv'(r)\bigg).\notag
\end{align}
Hence, the condition is satisfied for $i=1,j=2$ and the other cases follow similarly. 
The solution of \eqref{partlogf} is seen to be
\begin{equation}\label{fxyzsol}
f(x,y,z)=K_0 e^{-\frac{2V(r)}{\sigma^2}}
\end{equation}
and can be verified by direct differentiation. For example
\begin{align}\label{partlogfsol}
\frac{\partial}{\partial{x}}\log\left\{f(x,y,z)\right\}
&=-\frac{2}{\sigma^2}\frac{\partial}{\partial{x}} V(r)
=-\frac{2}{\sigma^2}v(r)\frac{\partial{r}}{\partial{x}} \notag \\
&=-\frac{2v(r)x}{\sigma^2 r}
=Z_1(x,y,z)
\end{align}
which satisfies\eqref{partlogf}. Similarly differentiation with respect to $y$ and $z$ verifies \eqref{fxyzsol} as the solution.

%

\section{Derivation of the 2D OU steady state distribution}\label{appenB}
Consider, for example, the SDEs for $(X_t,\epsilon_{1t})$, written in matrix form as:
		\begin{align}
		d\left[ \begin{array}{c} 
		X_t \\ \epsilon_{1t} 
		\end{array} \right] &=		- \begin{bmatrix} \alpha_1 & \alpha_1 \\ 0 & \beta_1 
		\end{bmatrix} 
		\left[ \begin{array}{c} X_t \\ \epsilon_{1t} \end{array} \right]dt 
		+ \begin{bmatrix} \sigma_1 & 0 \\ 0 & s_1 \end{bmatrix} 
		d\left[ \begin{array}{c} {W}_{1t} \\ B_{1t} \end{array} \right] \notag\\
				&=-A \left[ \begin{array}{c} 
		X_t \\ \epsilon_{1t} 
		\end{array} \right]    
		+Bd  
		\left[ \begin{array}{c} 
		W_{1t} \\ B_{1t} 
		\end{array} \right] 
				. \label{lambdaiderr}
		\end{align}
		From \cite[Sec. 4.5.6]{gardiner} the solution to \eqref{lambdaiderr} is Gaussian where thecovariancee matrix $\bm\Sigma$ of $(X_t,\epsilon_{1t})^T$ satisfies $\bm{A\Sigma}+\bm{\Sigma A}^T=\bm{BB}^T$. Solution of this gives \eqref{lambdai} as required.

\end{appendices}

\bibliographystyle{IEEEtran}
\bibliography{bibliographyUAV}

\end{document}